\documentclass[12pt]{article}
\usepackage{amsfonts}
\usepackage{graphicx}% Include figure files
\begin{document}

% \begin{titlepage}
\title{{\bf Level Excitation and Transition Probabilities of Some Nuclei in The Lower $fp$-Shell}}
\author{ Fouad Attia MAJEED$^{1,2}$  \\
 {\small $^1$Department of Physics, College of Science,}\\
 {\small Al-Nahrain University, Baghdad, Iraq}\\
 {\small and}\\
 {\small $^2$The Abdul
Salam International Centre for Theoretical Physics}\\
{\small fal-ajee@ictp.it}}
\vskip 1cm
\date{\today }
\maketitle
% \end{titlepage}

\begin{abstract}
Unrestricted shell model calculations in the lower $fp$-shell region
for the nuclei $^{46}$Ti, $^{46}$Cr and $^{46}$V have been performed
for the isovector T=1 positive parity states using the shell model
code OXBASH for Windows by employing the effective interactions
GXPF1, FPD6 and KB3G. The level schemes and transition strengths
$B$($E$2;$\downarrow$) are compared with the recently available
experimental data. A very good agreement were obtained for all
nuclei.
\end{abstract}

\section{\large Introduction}
The nuclear shell model has been very successful in our
understanding of nuclear structure: once a suitable effective
interaction is found, the shell model can predict various
observables accurately and systematically. For light nuclei, there
are several "standard" effective interactions such as the
Cohen-Kurath \cite{SC65} and the USD \cite{BA88} interactions for
the $p$ and $sd$ shells, respectively. On the other hand, in the
next major shell, $i.e.$ , in the $fp$-shell, there were also
"standard" interactions such as FPD6 \cite{WA91} and GXPF1
\cite{GX02}.

The spectroscopy of nuclei, in the $fp$-shell region, has been well
described within the shell model framework. Extensive shell model
calculations have been performed in this mass region, using several
model spaces and two-body interactions, the most remarkable work of
Brown and co-workers \cite{AG05, DC05, BF04, KL04, KY04, AF04,
SJ04}. Because of the quite importance of the $fp$-shell for variety
of problems in nuclear structure, such as electron capture in
supernova explosions. In this letter we report the shell model
calculations in the lower $fp$-shell region for the nuclei
$^{46}$Ti, $^{46}$Cr and $^{46}$V, to test the the ability of the
present effective interactions in reproducing the experiment in this
mass region.

\section{\large Shell model calculations}
\subsection{Excitation energies}
As mentioned in the earlier section, the main motivations for
studying these nuclei lies in the lower $fp$-shell due to the
importance of these in the recent applications in astrophysics and
because of the spin-orbit splitting that gives rise to a sizable
energy gap in the $pf$-shell between $f$$_{7/2}$ orbit and the other
orbits $p$$_{3/2}$ , $p$$_{1/2}$ and $f$$_{5/2}$, producing the N or
Z=28 magic number.

The calculations have been carried out using the code OXBASH
 for windows \cite{OX04} in the FP model space which comprised of the
$1p$$_{3/2}$, $1p$$_{1/2}$, $0f$$_{7/2}$ and $0f$$_{5/2}$ valence
orbits outside the $^{40}$Ca. Three effective interactions were
employed with FP model space for the calculations of level spectra
and transition probabilities, these effective in iterations are FPD6
\cite{WA91}, GXPF1 \cite{GX02} and KB3G \cite{AP01}. We should
mention here that $^{46}$Ti and $^{46}$Cr have only isovector part
T=1, while $^{46}$V have isovector part T=1 and isoscalar T=0, in
our study we considered only the isovector T=1 for $^{46}$V.

Figure 1. presents the comparison of the experimental excitation
energies of $^{46}$Ti with calculated values from FPD6, GXPF1 and
KB3G effective interactions. The three effective interactions gives
very good results in comparison with the experimental values up to
$J$$^{\pi}$=12$^{+}$. From Fig1. we can notice that FPD6 are in
excellent agreement with the experiment better than GXPF1 and KB3G.

\begin{figure}
\centering
\includegraphics[width=0.5\textwidth]{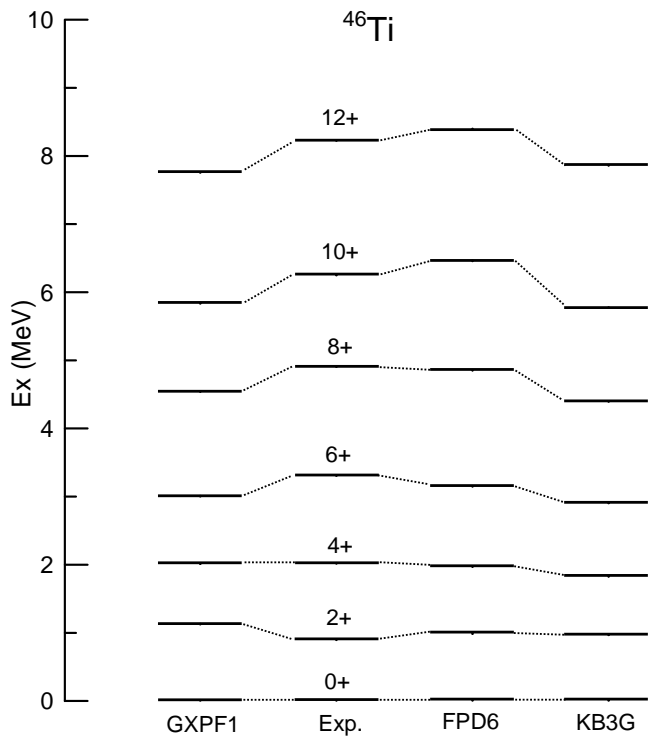}
\caption{\small Comparison of the experimental excitation energies
taken from Ref. \cite{PE01} with the present theoretical work using
FPD6, GXPF1 and KB3G effective interactions.}
\end{figure}

In figure 2 and figure 3, same comparison were made using the three
effective interactions for $^{46}$Cr and $^{46}$V respectively. From
these figures same conclusion were drawn that FPD6 is the best for
describing these nuclei lies in the lower part of the $fp$-shell.
\begin{figure}
\centering
\includegraphics[width=0.5\textwidth]{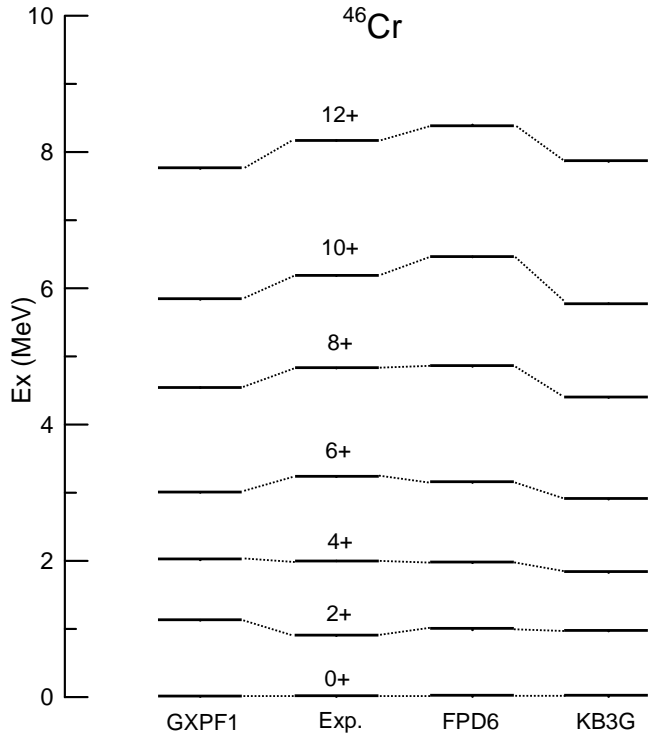}
\caption{\small Comparison of the experimental excitation energies
taken from Ref. \cite{PE01} with the present theoretical work using
FPD6, GXPF1 and KB3G effective interactions.}
\end{figure}

\begin{figure}
\centering
\includegraphics[width=0.5\textwidth]{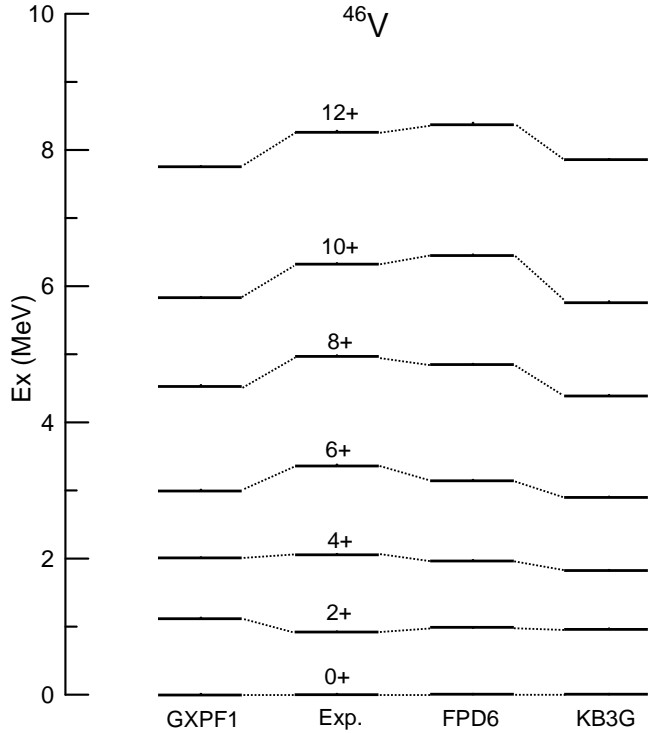}
\caption{\small Comparison of the experimental excitation energies
taken from Ref. \cite{FB01} with the present theoretical work using
FPD6, GXPF1 and KB3G effective interactions.}
\end{figure}

\subsection{Transition probabilities }
Since the transition rates represent a sensitive test for the most
modern effective interactions that have been developed to describe
$fp$-shell nuclei. The transition strengths calculated in this work
performed using the harmonic oscillator potential HO for each
in-band transition by assuming pure $E$2 transition. Core
polarization effect were included by choosing the effective charges
for proton $e$$_{\pi}$=0.7$e$ and for neutron $e$$_{\nu}$=0.5$e$.
Our results and the previous theoretical results using different
models are listed in Table 1 for $^{46}$Ti.

In Th.1 and Th.2 \cite{YH00},the effective charges for proton and
neutron were taken as 1.38$e$ and 0.83$e$ respectively. The
effective charges for protons and neutrons taken to be equal in
value as 0.7$e$ in Th.3 which is MONSTER \cite{KW84} and
$e$$_{\pi}$=$e$$_{\nu}$=0.9$e$ adopted in Th.4 "the
($f$$_{7/2}$)$^{6}$ shell model \cite{NR82}". As seen from Table 1,
the $B$($E$2;$\downarrow$) values calculated in this work are in
better agreement for the transitions $B$($E$2; 2$^+_1
\rightarrow$0$^+_1$) and $B$($E$2; 4$^+_1 \rightarrow$2$^+_1$) than
the previous theoretical work, while the rest transitions, Th1.,
Th.2, Th.3, Th.4 and Th.6 are in better agreement with the
experimental data, except Th.5 "the rotational model \cite{NR82}" do
not follow the trend of experimental data.

Although FPD6 effective interaction is more successful in
description of energy level spectra, but the calculation of the
transition strengths prove that it not the standard effective
interaction for this region and the results obtained by GXPF1 are in
excellent agreement with experiment, also the result of KB3G are not
so far from the experimental values.

For $^{46}$Cr the same comparison were made in Table 2, but the
experimental data are not available, therefore we can not judge
which effective interaction reproduce the experimental data better.

The effective charges for proton and neutron are taken to be 0.5$e$
and 0.4$e$ respectively, for the calculations of the transition
strengths of $^{46}$V. Our theoretical results are in excellent
agreement with the experimental values for the transitions $B$($E$2;
2$^+_1 \rightarrow$0$^+_1$) and $B$($E$2; 4$^+_1
\rightarrow$2$^+_1$) using GXPF1 effective interaction, also our
theoretical predictions are in better agreement from the previous
theoretical work Th.2 \cite{FB01} and Th.3 \cite{SM99} as summarized
in Table 3.

\begin{table}
\caption{\small The $B$($E$2) values in the ground-state band of
$^{46}$Ti. Their units are e$^{2}$ fm$^{4}$. Exp. is the experiment
\cite{NR82, LK93, SR87,FJ04}; Th.1 is PPNC; Th.2 is the projected of
the pure HF ground-state configuration \cite{YH00}; Th.3 is MONSTER
\cite{KW84}; Th.4 is the ($f$$_{7/2}$)$^{6}$ shell model
\cite{NR82}; Th.5 is the rotational model \cite{NR82}; Th.6 is
ANTOINE \cite{FJ04}. This work is assumed pure $E$2 transition
limit.}
\centering
\includegraphics[width=0.8\textwidth]{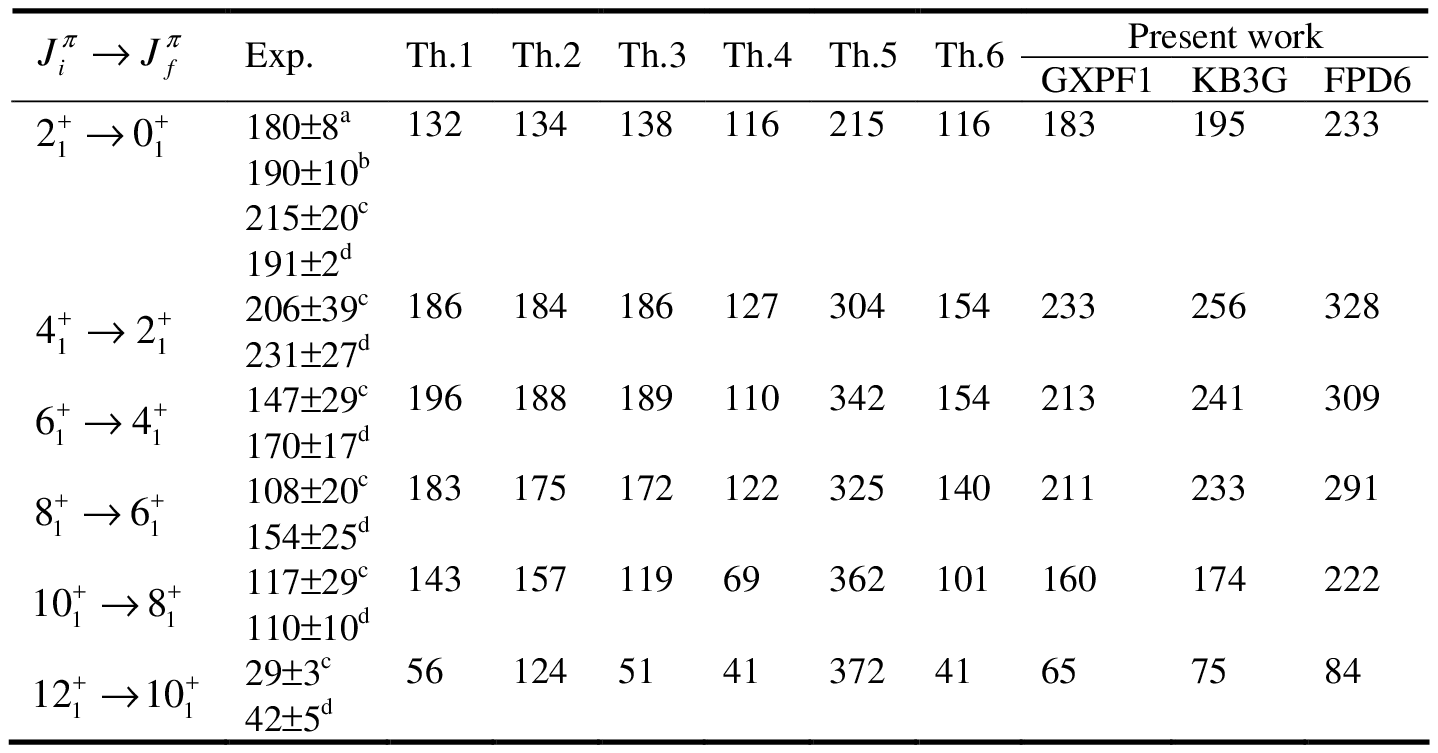}\\
\thanks {\small $^{a}$Reference\cite{LK93}, $^{b}$Reference\cite{SR87}, $^{c}$Reference\cite{NR82}, $^{d}$Reference\cite{FJ04}}
\end{table}

\begin{table}
\caption{\small The $B$($E$2) values in the ground-state band of
$^{46}$Cr. Their units are e$^{2}$ fm$^{4}$. Exp. is the experiment
\cite{KY05}. This work is assumed pure $E$2 transition limit.}
\centering
\includegraphics[width=0.5\textwidth]{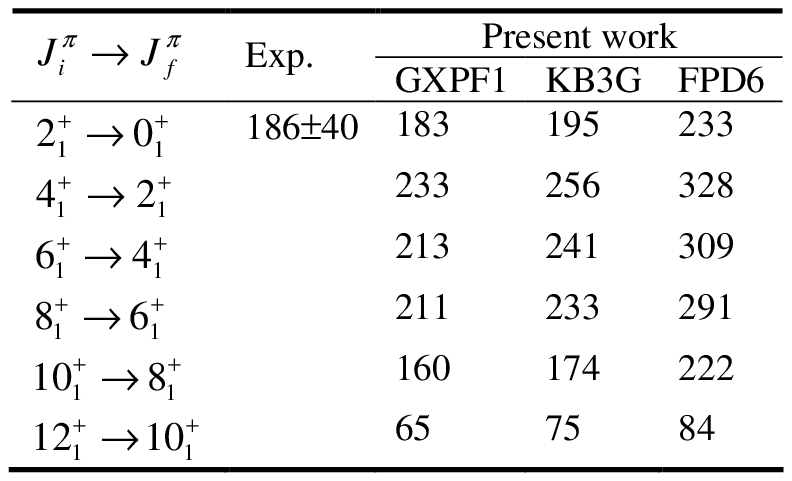}\\
\end{table}

\begin{table}
\caption{\small The $B$($E$2) values in the ground-state band of
$^{46}$V. Their units are e$^{2}$ fm$^{4}$. Exp. is the experiment
\cite{FB01, PB01}. This work is assumed pure $E$2 transition limit.}
\centering
\includegraphics[width=0.7\textwidth]{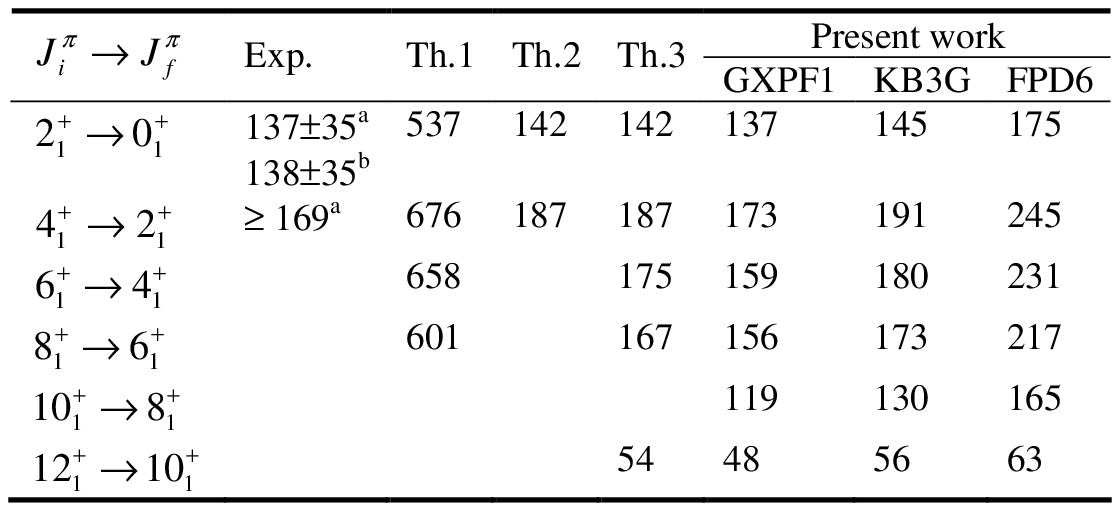}\\
\thanks {\small $^{a}$Reference\cite{PB01},$^{b}$Reference\cite{FB01}}

\end{table}

\newpage

\section{Summary}
Full $fp$-space shell model calculations were performed using the
code OXBASH for Windows. The FP model space were employed with the
effective interactions GXPF1, FPD6 and KB3G to reproduce the level
spectra and transition strengths $B$($E$2) for the nuclei $^{46}$Ti,
$^{46}$Cr and $^{46}$V. Excellent agreement were obtained by
comparing these calculations with the recently available
experimental data for the level spectra using FPD6 effective
interaction. Calculation of the transition strengths prove that
GXPF1 is more consistent in reproducing the experiment than FPD6 for
the lower $fp$-shell region.

\vskip 1cm

\noindent{\large{\bf Acknowledgement}}

\noindent This work is accomplished during my visit to the high
energy section of the Abdus Salam International Centre for
Theoretical Physics (ICTP). I would like to thank Prof. K. R.
Sreenivasan and Prof. S. Randjbar-Daemi for their kind invitation
and warm hospitality during my visit. Also I would like to
acknowledge the finical support from ICTP.

 %\newpage

\end{document}